# The Role of Employment Flexibility in Enhancing the Competitiveness of Temporary Staffing Service Providers in Poland

Michał Ćwiąkała[1], Dariusz Baran[2], Gabriela Wojak[3], Ernest Górka[4], Piotr Czarnecki[5], Patryk Paś[6], Marcin Kubera[7]

***Abstract:***

***Purpose:*** *This paper examines how employment flexibility, particularly through temporary staffing, enhances firms' competitiveness. It explores impacts on efficiency, cost reduction, workforce scalability, and client satisfaction. The study provides empirical evidence from Polish companies on the strategic role of flexible employment.*

***Design/methodology/approach:*** *A quantitative survey was conducted among managers of Polish firms using temporary staffing services. Purposeful sampling ensured responses from those with direct experience in flexible employment. The analysis focused on efficiency, cost, responsiveness, and satisfaction outcomes.*

***Findings:*** *Employment flexibility significantly reduces downtime, speeds up onboarding, and lowers personnel costs. It improves workforce scalability and access to specialized skills. Flexibility also increases client satisfaction and loyalty toward staffing service providers.*

***Practical implications:*** *The study is limited to Polish firms and relies on self-reported survey data. Future research should include qualitative methods and sector-specific analysis. Long-term effects and leadership roles in flexible employment require further exploration.*

***Originality/value:*** *Companies should integrate flexibility into strategic HR planning to enhance resilience and competitiveness. Investment in onboarding and close collaboration with agencies can improve workforce outcomes. Flexible staffing strategies help manage demand peaks and boost profitability.*

[1]*University College of Professional Education in Wroclaw, Poland, ORCID: 0000-0001-9706-864X, e-mail: michal.cwiakala@wsk.pl;*

[2]*Department of Social Sciences and Computer Science Nowy Sącz School of Business - National Louis University, Poland, ORCID:0009-0006-8697-5459; e-mail: dkbaran@wsb-nlu.edu.pl;*

[3]*I'M BRAND INSTITUTE Sp. z o.o., ORCID:0009-0003-2958-365X, e-mail: g.wojak@imbrandinstitute.com;*

[4]*The same as in 2, ORCID:0009-0006-3293-5670, e-mail: ewgorka@wsb-nlu.edu.pl;*

[5]*The same as in 2, ORCID:0009-0007-1041-8995, e-mail: pczarnecki@wsb-nlu.edu.pl;*

[6]*TIKROW Limited Liability Company, Poland, ORCID:0009-0003-5537-5440, e-mail: patryk.pas@tikrow.com;*

[7]*Pomeranian Higher School in Starogard Gdanski, Institute of Management, Economics and Logistics, Poland, ORCID: 0000-0002-6163-0650, e-mail: marcin.kubera@twojestudia.pl;*

1. Introduction

In an era defined by globalization, rapid technological advancement, and increasing labor market volatility, organizational adaptability has become a fundamental determinant of competitive advantage. Firms are increasingly required to respond swiftly and effectively to fluctuating market conditions, shifting consumer expectations, and evolving workforce dynamics.

Within this context, employment flexibility — particularly in the form of temporary staffing — has emerged as a strategic mechanism enabling organizations to maintain operational continuity, reduce labor costs, and sustain long-term competitiveness (Blunsdon, 2017; Sakowska, 2025).

As companies seek to navigate an environment characterized by uncertainty and disruption, workforce flexibility is no longer viewed merely as a tactical adjustment but as a core strategic capability underpinning organizational resilience and sustainable growth (Taracha and Mirowski, 2022; Azizzadeh *et al.,* 2022; Islam *et al.,* 2023; Zupok, 2009).

The extant literature has examined employment flexibility primarily from macroeconomic and social perspectives, addressing issues such as its influence on labor market dynamics, job quality, and employee well-being (Aleksynska, 2018; Lisi, 2017). However, considerably less attention has been devoted to the strategic implications of flexible employment for enterprise-level competitiveness, particularly in relation to organizations that rely on temporary staffing services.

Existing studies often overlook how specific dimensions of flexibility — including workforce scalability, accelerated recruitment processes, cost optimization, and enhanced market responsiveness — translate into tangible performance outcomes and strategic advantages (Feliczek, 2024; Gołaś and Organiściak, 2019).

This gap is especially pertinent in the Polish context, where the temporary staffing sector has expanded significantly yet remains underexplored in comparison with more mature markets in Western Europe (Polskie Forum HR, 2024).

The present study addresses this research gap by adopting an empirical approach that foregrounds the client-company perspective rather than the more commonly investigated macroeconomic or employee-centric dimensions.

By integrating classical theoretical frameworks — such as Atkinson's "flexible firm" model (1970s) and the resource-based view of human capital (Matwiejczuk, 2015), with contemporary developments in digital transformation and the platformization of human resource management (Panek, 2024; Peicheva, 2022), this research contributes to a deeper understanding of the strategic role of employment flexibility. Specifically, it examines how flexible workforce arrangements enhance organizational adaptability, operational efficiency, and competitive positioning in increasingly dynamic business environments.

## 2. Literature Review

Employment flexibility has been conceptualized in various ways in the literature, reflecting both different research perspectives and sector-specific conditions. Quantitative flexibility emphasizes the organization's ability to rapidly adjust workforce size by hiring or reducing employees in response to fluctuating demand, while qualitative flexibility refers to workforce versatility and retraining, enabling job rotation within organizational structures (Gołaszewska-Kaczan, 2013).

Another dimension, temporal flexibility, focuses on adjusting working hours through flexible schedules or task-based systems, whereas functional flexibility emphasizes adaptive use of employees' skills to dynamically reallocate roles within the organization.

These approaches collectively enable companies to react promptly to changing environmental conditions by integrating numerical, functional, temporal, and financial dimensions (Blunsdon, 2017).

Later studies introduced the spatial dimension, addressing remote and hybrid work models as a response to geographic dispersion and digitalization trends (Sidor-Rządkowska, 2022).

Atkinson's (1970s) "flexible firm" model established four core dimensions of organizational adaptability: quantitative, temporal, functional, and financial. His work underscored that rigid workforce structures reduce firms' resilience to economic fluctuations (Leśniewski, 2023).

Similarly, contingency theory highlighted that dynamic and unpredictable environments demand flexible human resource management practices (Pędziwiatr, 2018). The resource-based view, emerging in the 1990s, added that human capital represents a strategic asset only if organizations can redeploy and develop it in a flexible manner (Matwiejczuk, 2015).

Contemporary research extends these frameworks by incorporating globalization, digitalization, and platform-based work models. Remote work, geographically dispersed teams, and hybrid organizational structures illustrate how spatial flexibility complements traditional dimensions, enhancing managerial capacity to respond to volatility and innovation pressures (Sakowska, 2025).

Consequently, employment flexibility is no longer solely a cost-reduction mechanism but a strategic capability underpinning organizational resilience and competitiveness in dynamic markets (Taracha and Mirowski, 2022).

Temporary work represents a specific form of employment governed by a tripartite relationship involving the temporary work agency, the temporary worker, and the user employer. The agency, as the formal employer, signs a contract with the worker, delegating them to the user employer who supervises work tasks and workplace conditions (Dudek, 2011).

Legal frameworks, such as the Polish Act on the Employment of Temporary Workers (2003) and Directive 2008/104/EC, set strict limits: a temporary worker may be assigned to the same user employer for no longer than 18 months within a 36-month period, ensuring employee protection and preventing misuse of temporary contracts (Official Journal EU, 2008).

Temporary work arrangements vary by purpose and duration. Seasonal employment addresses cyclical demand fluctuations in sectors such as tourism and agriculture; replacement contracts cover prolonged staff absences; project-based employment supports time-bound initiatives requiring specialized skills; and on-demand models leverage digital platforms for rapid workforce mobilization in response to unforeseen operational needs (Czapliński, 2010; Pańkow, 2014).

Temporary staffing agencies play a dual role: operationally, they minimize transaction costs related to recruitment, selection, and training (Florek, 2021), while strategically, they advise firms on workforce planning, wage flexibility, and risk management (Rogozińska-Pawelczyk, 2014). Modern agencies employ advanced tools such as psychometric testing, assessment centers, and predictive analytics to enhance candidate-job fit, thereby increasing efficiency and reducing turnover risk (Snopkiewicz, 2011).

Market data indicate steady growth in Poland's temporary staffing sector, with temporary workers constituting 1.5% of total employment in 2023 compared to 4–5% in countries like the Netherlands or Germany, where flexible labor regulations and outsourcing cultures are more developed (Polskie Forum HR, 2024).

Globally, the temporary labor market grows by 5–7% annually, driven by e-commerce, logistics, and IT sectors, as well as the rapid adoption of on-demand platforms accelerated by the COVID-19 pandemic (Grand View Research, 2025).

The decision to employ temporary workers stems from both external market pressures and internal organizational strategies. Cyclical demand fluctuations - such as peak seasons in retail or logistics - necessitate rapid workforce scaling to avoid overtime costs or production bottlenecks (de Graaf Zijl and Berkhout, 2007). Conversely, during downturns, temporary contracts enable cost reduction without lengthy redundancy procedures, stabilizing operational margins (Schultz, 2023).

Financial flexibility is another critical factor. By converting fixed labor costs - wages, benefits, and insurance - into variable expenses tied to actual working hours, firms improve liquidity and reduce budgetary risk (Mędrala, 2022). Cost simulations comparing full-time employment with partial outsourcing demonstrate that temporary staffing mitigates financial exposure while preserving workforce scalability (Gołaś and Organiściak, 2019).

Moreover, temporary staffing addresses talent shortages in specialized sectors like advanced manufacturing or IT. Agencies maintain extensive candidate networks and employ predictive analytics to match skills with job requirements efficiently, reducing hiring lead times by up to 50% compared to traditional recruitment methods (Feliczek, 2024).

Regulatory frameworks impose additional constraints, including contract duration limits, minimum wage requirements, and sector-specific certifications (Krzyśków and Milewska, 2009). Compliance with health, safety, and labor standards necessitates close cooperation between legal, HR, and operational departments, ensuring that flexibility does not compromise worker protection or corporate reputation (Bąk-Grabowska, 2013).

Finally, digital transformation reshapes temporary staffing through applicant tracking systems (ATS), artificial intelligence–driven candidate preselection, and on-demand platforms integrating real-time job matching with enterprise resource planning (ERP) systems. These technologies accelerate recruitment cycles, enhance talent profiling accuracy, and enable predictive workforce planning, strengthening temporary staffing as a strategic HR tool (Peicheva, 2022; Panek, 2024; Ostoj, 2024).

Recent research indicates that the effectiveness of flexible employment practices, including temporary work and variable scheduling, depends not only on organizational structures but also on leadership quality. Kasperczuk *et al.* (2025) and Ćwiąkała *et al.* (2025) highlight that leaders' emotional intelligence, empathy, and ethical conduct are strongly correlated with team motivation and engagement.

Employees perceive leaders who manage emotions effectively and act ethically as more trustworthy and supportive, which enhances commitment and reduces turnover in flexible work arrangements. Consequently, organizations combining flexible HR practices with the development of leaders' interpersonal competencies are better positioned to maintain employee motivation, optimize performance, and ensure stability in dynamic labor markets.

Temporary employment, characterized by short-term contracts, offers both advantages and challenges for workers and employers alike. On the positive side, it provides flexibility, allowing workers to gain diverse experience and employers to adjust staffing levels according to demand (Mackinnon and Partners, 2025).

However, numerous studies highlight significant drawbacks. Temporary workers often face lower job satisfaction, limited access to training, and reduced opportunities for career advancement compared to their permanent counterparts (Aleksynska, 2018). Research also indicates that temporary employment can have a negative impact on productivity growth, particularly in skilled sectors (Lisi, 2017).

Despite these challenges, temporary positions can serve as stepping stones to permanent employment, provided there are opportunities for skill development and organizational support (De Cuyper *at al.,* (2011).

### 3. Research Methodology and Case Description

In order to assess the impact of flexible forms of employment on the competitiveness of firms using temporary staffing services, a quantitative study was conducted using a structured survey among managers and owners of client companies cooperating with temporary work agencies. Purposeful sampling was applied, selecting respondents who had direct experience with flexible employment models, which allowed the collection of informed opinions from individuals actively involved in personnel management decisions.

Temporary staffing agencies and short-term employment contracts represent flexible labor solutions that enable companies to respond dynamically to market demands, reduce downtime, and optimize operational efficiency. The study focused on three key dimensions of operational effectiveness: reduction of downtime, acceleration of onboarding processes, and the ability to scale workforce resources during periods of increased demand.

The choice of firms using temporary staffing services as the study context was based on several considerations:

- ➢ The labor market in Poland has seen increasing reliance on flexible employment, particularly in sectors subject to seasonal fluctuations and variable demand. Understanding the effects of such flexibility is essential for designing competitive workforce strategies.
- ➢ There is limited empirical research analyzing the specific impact of temporary staffing and short-term contracts on operational performance and firm competitiveness. Existing literature often discusses flexibility in general terms without assessing measurable outcomes in the context of client-company operations.

➢ From a practical standpoint, efficiency in resource allocation, rapid adaptation to market changes, and cost optimization are critical determinants of firm competitiveness. Flexible employment, by allowing scalable human resource deployment, can contribute directly to operational resilience and client satisfaction.

The research was limited to Polish firms that had used at least one temporary staffing service in the past five years, excluding entities relying solely on permanent employment.

The study involved 92 respondents from diverse sectors. Services represented the largest group (32.6%, n = 30), followed by production (26.1%, n = 24), transport and logistics (19.6%, n = 18), and trade (15.2%, n = 14). Other sectors accounted for 6.5% (n = 6), illustrating the presence of niche applications of temporary employment.

Company size distribution indicated a predominance of medium-sized firms (43.5%, n = 40, employing 10–49 employees), followed by firms with 50–249 employees (32.6%, n = 30). Micro-entities (1–9 employees) and larger firms (250–999 employees) both represented 10.9% (n = 10), while very large companies (≥1000 employees) were rare (2.2%, n = 2).

In terms of annual turnover, 37.0% of respondents represented companies with revenues of 1–10 million PLN (n = 34), 30.4% had 10–50 million PLN (n = 28), smaller companies (<1 million PLN) accounted for 8.7% (n = 8), medium-large enterprises (50–200 million PLN) represented 17.4% (n = 16), and the largest (>200 million PLN) 6.5% (n = 6).

The duration of cooperation with agencies showed that 32.6% (n = 30) of firms had collaborated for 5–10 years, 28.3% (n = 26) for 3–5 years, 21.7% (n = 20) for 1–3 years, 13.0% (n = 12) for over a decade, and 4.3% (n = 4) for less than one year.

The average number of temporary staff employed annually was 11–50 employees in 45.7% of firms (n = 42), 51–100 in 23.9% (n = 22), 1–10 in 17.4% (n = 16), 101–500 in 10.9% (n = 10), and over 500 employees in 2.2% (n = 2).

The sample demonstrates that temporary staffing is particularly significant for medium-sized service and production companies, with budgets ranging mostly from 1–50 million PLN and 3–10 years of agency collaboration, which provides a solid basis for drawing conclusions relevant to the SME sector.

Data collection occurred between May and June 2025 using a hybrid distribution model: paper questionnaires delivered to company offices and an online Google

Forms survey disseminated via professional LinkedIn connections. This approach ensured a representative cross-section of managerial opinions, yielding 92 completed surveys for quantitative analysis.

### 4. Research Results

The first part of the analysis focused on the impact of flexible forms of employment on the operational efficiency of companies using temporary work services. The respondents were asked to evaluate several statements related to downtime reduction, onboarding speed, overtime levels, internal process fluidity, and workforce scalability, using a 5-point Likert scale (1 – strongly disagree; 5 – strongly agree).

*Table 1. Assessment of the impact of employment flexibility on operational efficiency*

| Item | Mean | Median | SD |
|---|---|---|---|
| Downtime decreased thanks to flexible employment | 4,08 | 4,0 | 0,99 |
| Temps were onboarded faster than permanent staff | 3,87 | 4,0 | 1,08 |
| Temps reduced overtime among core staff | 3,34 | 3,0 | 1,15 |

**Source:** *Own elaboration based on conducted research.*

The results show that flexible employment contributes significantly to process optimization. The majority of respondents agreed that the introduction of temporary employment solutions reduced downtime in task execution (M = 4,08; SD = 0,99) and facilitated faster integration of temporary employees into organizational structures compared to permanent staff (M = 3.87; SD = 1,08). Moreover, a significant share of participants confirmed that the use of temporary employees reduced the overtime burden on permanent staff (M = 3,34; SD = 1,15).

*Table 2. Distribution of responses on the impact of flexible employment on operational efficiency*

| Item | 1 (n) | 2 (n) | 3 (n) | 4 (n) | 5 (n) |
|---|---|---|---|---|---|
| Flexibility improved internal process flow | 4 | 11 | 10 | 40 | 37 |
| Faster scaling of workforce during peaks | 1 | 2 | 14 | 30 | 45 |

**Source:** *Own elaboration based on conducted research.*

These results indicate that flexibility plays a critical role in increasing organizational adaptability. Over 83% of companies reported that flexible employment improved the continuity of internal processes, and more than 81% highlighted its role in enabling rapid workforce expansion during demand peaks.

The second area of the study investigated the extent to which flexible employment models influence personnel costs, recruitment expenditures, and overall profitability.

The analysis shows that the majority of organizations perceive significant financial benefits from employing temporary workers.

***Table 3.*** *Perceived impact of employment flexibility on cost reduction and profitability*

| Item | Agree 4-5 (n) | Neutral 3 (n) | Disagree 1-2 (n) |
|---|---|---|---|
| Lower total personnel costs | 62 | 15 | 15 |
| Improved margin on services/products | 59 | 19 | 14 |
| Lower recruitment/selection costs vs permanent | 63 | 17 | 12 |
| Avoided leave-related expenditures | 52 | 21 | 19 |
| Higher profitability of short-term projects | 65 | 16 | 11 |

**Source:** *Own elaboration based on conducted research.*

Nearly 68% of respondents agreed that cooperation with temporary employment agencies reduces total personnel costs. Approximately two-thirds confirmed that flexibility improves profit margins on products and services, while 68.5% indicated that recruitment and selection costs for temporary workers are lower than those for permanent staff. Furthermore, more than 70% of participants reported increased profitability in short-term projects resulting from flexible workforce deployment.

The survey also explored how flexible forms of employment affect workforce quality and access to specialized skills. The results indicate that companies highly value the qualifications of temporary workers, with three-quarters of respondents stating that they possessed the skills necessary for their roles. Additionally, most respondents rated their performance quality as comparable to that of permanent employees.

While fewer than half of the organizations considered pre-employment training to be highly effective, many emphasized its importance in enhancing readiness for work. Approximately 70% of respondents reported that flexible employment enabled them to quickly access experts in specialized fields without lengthy recruitment processes. Most also confirmed that temporary employees rapidly adapted to internal procedures, allowing them to achieve full productivity in a shorter time.

Another critical dimension of employment flexibility is its impact on organizational responsiveness to changing market conditions. The research clearly demonstrates that flexible workforce models enhance the ability of companies to adapt to fluctuations in demand and market dynamics.

***Table 4.*** *Assessment of employment flexibility in enhancing market responsiveness*

| Item | Agree 4-5 (n) | Neutral 3 (n) | Disagree 1-2 (n) |
|---|---|---|---|
| Rapid staff increase during demand spikes | 72 | 12 | 8 |
| Rapid staff reduction in slowdowns | 61 | 17 | 14 |
| Faster entry to new markets | 56 | 19 | 17 |
| Short-term project execution without long-term commitments | 65 | 17 | 10 |
| Greater resilience to seasonal fluctuations | 72 | 12 | 8 |

**Source:** *Own elaboration based on conducted research.*

More than 78% of respondents confirmed that flexible employment allowed them to rapidly scale their workforce during periods of increased demand, while over 66% reported that it enabled swift employment reduction during market slowdowns. Additionally, around 61% indicated that flexibility facilitated faster market entry, and more than 70% acknowledged its role in executing short-term projects without long-term obligations. Almost the same proportion emphasized that employment flexibility increased their resilience to seasonal variations.

The final part of the study examined how flexible employment influences client satisfaction and loyalty toward temporary employment service providers. The data clearly indicate that flexibility is a key determinant of positive client relations.

*Table 5. Evaluation of client satisfaction and loyalty factors*

| Item | Agree 4-5 (n) | Neutral 3 (n) | Disagree 1-2 (n) |
|---|---|---|---|
| Satisfaction with agency service quality | 82 | 6 | 4 |
| Flexibility of offer increases satisfaction | 72 | 11 | 9 |
| Willingness to recommend the agency | 60 | 19 | 13 |
| Loyalty: returning to the same agency | 56 | 17 | 19 |
| Flexibility as a decisive selection criterion | 69 | 11 | 12 |

**Source:** *Own elaboration based on conducted research.*

Almost 89% of respondents expressed satisfaction with the quality of services received, while 78% emphasized that flexibility was a crucial factor in their overall satisfaction. More than 65% of clients declared their willingness to recommend their current service provider to other companies, and over 60% reported consistent loyalty to a single agency. Furthermore, 75% of respondents stated that employment flexibility was a decisive factor in selecting a partner for workforce services.

## 5. Conclusions and Future Research Implications

The conducted study provides robust empirical confirmation that employment flexibility, particularly through temporary staffing services, constitutes a strategic determinant of organizational competitiveness. Building on the theoretical frameworks outlined in the literature review – including Atkinson's flexible firm model (1970s), the contingency approach to HRM (Pędziwiatr, 2018), and the resource-based view of human capital (Matwiejczuk, 2015) – the findings demonstrate that flexible employment is not limited to cost optimization but extends to resilience, adaptability, and client satisfaction.

From an operational perspective, the results support earlier research highlighting the significance of numerical and temporal flexibility in mitigating downtime and ensuring continuity of processes (Gołaszewska-Kaczan, 2013; Blunsdon, 2017). Over four-fifths of surveyed companies confirmed that temporary employment enables rapid scaling of workforce resources, validating the notion that flexible forms of work are central to maintaining efficiency in volatile environments.

The research also underscores the importance of qualitative and functional flexibility by showing that temporary employees are generally perceived as competent and capable of rapid adaptation. This aligns with studies stressing that the strategic value of flexibility emerges when it is combined with the development and deployment of human capital (Feliczek, 2024; Kasperczuk *et al.,* 2025). The ability to access specialists on short notice without protracted recruitment processes confirms the resource-based view that flexibility amplifies the strategic role of human capital.

Finally, the survey results confirm that client satisfaction and loyalty are strongly shaped by flexible staffing solutions. Nearly 90% of respondents expressed satisfaction with the quality of services, and more than three-quarters viewed flexibility as a decisive selection criterion for agencies.

The study suggests several managerial recommendations:

- Integrate flexibility into strategic HR planning, treating it as a resilience mechanism rather than a purely reactive tool.
- Leverage temporary staffing for peak-demand management, reducing overtime costs and ensuring timely task completion.
- Invest in onboarding and pre-employment training systems to maximize the effectiveness of temporary staff from the first day of work.
- Strengthen cooperation with agencies through shared forecasting and transparent communication to improve candidate availability and alignment with company needs.

This research is subject to certain limitations. The reliance on a quantitative survey of 92 Polish firms constrains the generalizability of results to broader contexts. The study also did not differentiate outcomes by sectoral volatility, leadership style, or the specific digital tools used by staffing agencies.

Future studies should address these gaps by:

- Combining quantitative and qualitative approaches (e.g., in-depth interviews, case studies) to capture managerial experiences and worker perspectives in greater depth.
- Comparing sector-specific effects of employment flexibility, particularly in industries with varying degrees of seasonality and specialization.

- ➢ Exploring digital and spatial dimensions of flexibility, including the role of remote and hybrid models in enhancing competitiveness.
- ➢ Assessing long-term outcomes of flexible employment on employee career development, job satisfaction, and organizational productivity, building on concerns raised in earlier studies (Aleksynska, 2018; Lisi, 2017).
- ➢ Investigating leadership and organizational culture as moderators of flexibility outcomes, given the evidence linking managerial competencies to motivation and engagement in flexible settings (Ćwiąkała *et al.,* 2025).

In conclusion, the study confirms that employment flexibility is a multidimensional strategic asset. It improves operational efficiency, reduces costs, enhances resilience, and strengthens client relationships. When aligned with broader organizational strategies and supported by competent leadership, flexibility transforms from a tactical adjustment tool into a driver of sustainable competitiveness in dynamic labor markets.